\documentclass[titlepage,floatfix,floats,superscriptaddress,reprint,prl,aps]{revtex4-1}

\usepackage{times}
\usepackage{graphicx}
\usepackage{amsmath}
\usepackage{amssymb}
\usepackage{amsbsy}
\begin{document}

\title{A Variational Approach to Enhanced Sampling and Free Energy Calculations}
\author{Omar Valsson}
\email{omar.valsson@phys.chem.ethz.ch}
\author{Michele Parrinello}
\email{parrinello@phys.chem.ethz.ch}
\affiliation{Department of Chemistry and Applied Biosciences, ETH Zurich and 
            Facolt\`{a} di Informatica, Instituto di Scienze Computationali, 
             Universit\`{a} della Svizzera italiana, 
             Via Giuseppe Buffi 13, CH-6900, Lugano, Switzerland.}

\date{August 5, 2014}

\begin{abstract}

The ability of widely used sampling methods, such as  molecular dynamics or 
Monte Carlo,  to explore complex free energy landscapes is severely hampered by 
the presence of kinetic bottlenecks. A large number of  solutions have been 
proposed to alleviate this problem.  Many are based on the introduction of a 
bias potential which is a function of a small number of collective variable.  
However constructing such a  bias is not simple.  Here we introduce a  
functional of the bias potential and an associated variational principle. The 
bias that  minimizes the functional  relates in a simple way to the free energy 
surface. This variational principle can be turned into a practical, efficient and 
flexible sampling method. A number of numerical examples are presented which 
include the determination of a three dimensional free energy surface. We argue 
that, beside being numerically advantageous, our variational approach provides a 
convenient standpoint for looking with novel eyes at the sampling problem.

\end{abstract}

\maketitle


Molecular Dynamics (MD) and Monte Carlo (MC) simulations have become 
indispensable tools in many areas of science. However, whenever there are 
kinetic bottlenecks that lead to  the  appearance of long-lived metastable 
states, the computational cost of sampling the systems configuration space 
becomes prohibitive. 
This has lead to an intensive search for enhanced methods 
capable of lifting this severe limitation.
One of the oldest such methods and still much in use today is umbrella 
sampling~\cite{Torrie1977a} in which an external bias is added to the system to 
favor transitions between states separated by kinetic barriers and allows them 
to occur on the time scale of the simulation. However building such potential is 
very challenging and many methods have been devised to this effect
~\cite{Huber1994,Darve2001,WangLandau2001a,Laio2002a,Hansmann2002a,
Maragliano2006a,Abrams2010a,Maragakis2009a}.

Here we present a new and efficient approach to this problem and propose a 
variational method that allows constructing an effective bias potential and 
leads to an accurate determination of the free energy as a function of a set of 
chosen collective variables (CVs). 

In the following we consider a system described by the  microscopic coordinates 
$\mathbf{R} \in \mathbb{R}^{3N}$ whose dynamics (e.g.\ MD or MC) at temperature 
$T$ evolves according to a potential energy function $U(\mathbf{R})$, and leads 
to a canonical equilibrium distribution 
$P(\mathbf{R})= e^{-\beta U(\mathbf{R})}/Z$ 
where $\beta = (k_{\mathrm{B}} T)^{-1}$ is the inverse temperature 
and $Z = \int d\mathbf{R} \, e^{-\beta U(\mathbf{R})}$ 
is the partition function of the system. 
We map the high dimensional $\mathbf{R}$ space into a much smaller and smoother 
$d$ dimensional space by introducing the set of collective variables 
$\mathbf{s}(\mathbf{R}) = \left(s_{1}(\mathbf{R}),s_{2}(\mathbf{R}), \ldots, 
s_{d}(\mathbf{R})\right)$ 
that give a coarse grained but physically cogent description of the system. 
The appropriate choice of these collective variables is much discussed in the 
literature~\cite{Barducci2011a} and here we assume that their selection  has 
been wise.
The free energy surface (FES) associated to the CV set $\mathbf{s}$ is defined 
up to constant as 
\begin{equation}
\label{fes1}
F(\mathbf{s}) = - \frac{1}{\beta} \log
\int d\mathbf{R} \, \delta (\mathbf{s} - \mathbf{s}(\mathbf{R}) ) e^{-\beta 
U(\mathbf{R})}.
\end{equation}
The corresponding equilibrium distribution is 
$P(\mathbf{s}) = e^{-\beta F(\mathbf{s})}/Z$ and the partition function can be 
rewritten  as $Z = \int d\mathbf{s} \, e^{-\beta F(\mathbf{s})}$.

We introduce now  the following functional of a bias potential $V(\mathbf{s})$ 
\begin{align}
\label{omega1}
\Omega [V] & = 
\frac{1}{\beta} \log
\frac
{\int d\mathbf{s} \, e^{-\beta \left[ F(\mathbf{s}) + V(\mathbf{s})\right]}} 
{\int d\mathbf{s} \, e^{-\beta F(\mathbf{s})}} 
+ 
\int d\mathbf{s} \, p(\mathbf{s}) V(\mathbf{s})
\end{align}
where $p(\mathbf{s})$  is arbitrary probability distribution that is assumed to 
be normalized. 
The second term can thus be read  as the expectation value of $V(\mathbf{s})$ 
over the distribution $p(\mathbf{s})$.
As shown in the Supplemental Material (SM), this functional is convex and  
invariant under the addition of an arbitrary constant 
to $V(\mathbf{s})$, $\Omega [V+k] = \Omega [V]$ .

The potential that  renders 
 $\Omega [V]$ stationary is, within an irrelevant constant,
\begin{equation}
\label{optimal_bias}
V(\mathbf{s}) = -F(\mathbf{s})-{\frac {1}{\beta}} \log {p(\mathbf{s})}
\end{equation}
for   $p(\mathbf{s}) \neq 0$ 
and $V(\mathbf{s})=\infty$ otherwise.
This stationary point is also the  global minimum of $\Omega[V]$ since the 
functional is convex.
When the optimal bias potential (Eq.~\ref{optimal_bias}) acts 
on the system the $\mathbf{s}$ values sampled will be 
only those for which $p(\mathbf{s}) \neq 0$  and   $p(\mathbf{s})$ will be their  
resulting distribution.
This offers the interesting possibility of selecting the region in CV space to 
be explored by appropriately choosing $p(\mathbf{s})$ (see below and SM).
In more general terms the freedom of choosing  $p(\mathbf{s})$ confers  a high 
degree of flexibility to the method. 

If the CVs are defined in a compact phase space of volume $\Omega_{\mathbf{s}}$ 
a one possible and perhaps natural 
choice is to take $p(\mathbf{s})=\frac {1} {\Omega_{\mathbf{s}}}$ 
which leads to a uniform sampling in CV space 
as commonly done in other enhanced sampling approaches.
In this case 
$V(\mathbf{s})=-F(\mathbf{s})$ modulo a constant which is the same relation as 
is obeyed in the asymptotic limit by the bias potential in standard 
metadynamics~\cite{Laio2002a}.
If the CVs are unbound then $p(\mathbf{s})$ can be employed to focus on  the 
range of $\mathbf{s}$ to be explored.
Another  possibility is to use as $p(\mathbf{s})$ 
\begin{equation}
p(\mathbf{s})=\frac{e^{-\beta' F(\mathbf{s})}}{\int d\mathbf{s} \, e^{-\beta' 
F(\mathbf{s})}}
\end{equation} 
where  $\beta'= (k_{\mathrm{B}} (T+\Delta T))^{-1}$ 
and $F(\mathbf{s})$ is our target free energy surface at inverse temperature 
$\beta$ 
as defined in Eq.~\ref{fes1} above.
This is the distribution sampled in well-tempered metadynamics with a bias 
factor $\gamma=\frac {\beta}{\beta'}$~\cite{Barducci2008_WTMetaD}. 
With this choice the relation between the bias potential 
and free energy becomes identical to the one that asymptotically holds in 
well-tempered 
metadynamics~\cite{Barducci2008_WTMetaD,Dama2014a}, 
$V(\mathbf{s})=-(1-\frac {1}{\gamma} )F(\mathbf{s})$. We shall defer to a future 
publication the exploration 
of this intriguing possibility. 
Finally we note that it is our belief that an appropriate choice of $p(\mathbf{s})$ and smart usage of the variational flexibility of the bias potential can be of great help when considering difficult multidimensional CV spaces. We intend to explore this further in the future.

To make use of the variational  property of  $\Omega [V]$  we  write the bias 
potential $V(\mathbf{s};\boldsymbol{\alpha})$ as a function of the set of 
variational parameters 
$\boldsymbol{\alpha}= \left(\alpha_{1},\alpha_{2},\ldots,\alpha_{K}\right)$ 
and then minimize the function 
$\Omega(\boldsymbol{\alpha}) = \Omega[V(\boldsymbol{\alpha})]$ with 
respect to $\boldsymbol{\alpha}$. Of course the search for the minimum will be 
greatly facilitated  by the convexity of the functional.
From the converged potential $V(\mathbf{s};\boldsymbol{\alpha})$ we can then 
estimate $F(\mathbf{s})$ directly from Eq.~\ref{optimal_bias} if the assumed 
functional form has enough variational flexibility.
Otherwise, we can always estimate the FES as a function of $\mathbf{s}$, or some 
other CVs, by employing the standard umbrella sampling relation
\begin{equation}
\label{US_reweighting}
P(\mathbf{R}) \propto e^{\beta V(\mathbf{s}(\mathbf{R}))} P_{V}(\mathbf{R}) 
\end{equation}
where $P_{V}(\mathbf{R})$ is the  distribution biased by
$V(\mathbf{s}(\mathbf{R}))$ (see the SM for further discussion on 
this equation). 
The reweighting can also be performed before the potential has fully converged 
or even on the fly during the optimisation if the biasing potential converges 
quickly to a quasi-stationary state during the optimisation process.

In order to implement the optimisation procedure, we shall need to 
estimate the gradient 
$\Omega'(\boldsymbol{\alpha})$, 
\begin{equation}
\label{gradient1}
\frac
{\partial \Omega(\boldsymbol{\alpha})}
{\partial \alpha_{i}}
= - 
\left< 
\frac
{\partial V(\mathbf{s};\boldsymbol{\alpha})}
{\partial \alpha_{i}}
\right>_{V(\boldsymbol{\alpha})} +
\left< 
\frac
{\partial V(\mathbf{s};\boldsymbol{\alpha})}
{\partial \alpha_{i}}
\right>_{p},
\end{equation}
and the Hessian $\Omega''(\boldsymbol{\alpha})$, 
\begin{align}
\label{hessian1}
\frac
{\partial^{2} \Omega(\boldsymbol{\alpha})}
{\partial \alpha_{j} \partial \alpha_{i}}
& = \beta \cdot 
\mathrm{Cov}
\left[
\frac
{\partial V(\mathbf{s};\boldsymbol{\alpha})}
{\partial \alpha_{j}},
\frac
{\partial V(\mathbf{s};\boldsymbol{\alpha})}
{\partial \alpha_{i}}
\right]_{V(\boldsymbol{\alpha})}
\nonumber 
\\ 
 & - 
\left< 
\frac
{\partial^{2} V(\mathbf{s};\boldsymbol{\alpha})}
{\partial \alpha_{j} \partial \alpha_{i}}
\right>_{V(\boldsymbol{\alpha})} +
\left< 
\frac
{\partial^{2} V(\mathbf{s};\boldsymbol{\alpha})}
{\partial \alpha_{j} \partial \alpha_{i}}
\right>_{p},
\end{align}
where $\left< \cdots \right>_{V(\boldsymbol{\alpha})}$ and 
$\mathrm{Cov}[\cdots]_{V(\boldsymbol{\alpha})}$ are the 
the expectation value and the covariance, respectively,  
obtained in a biased simulation employing the 
potential $V(\mathbf{s};\boldsymbol{\alpha})$ 
and $\left< \cdots \right>_{p}$ is an expectation value 
in the distribution $p(\mathbf{s})$.
A natural approach is  to expand $V(\mathbf{s};\boldsymbol{\alpha})$ 
in a linear basis set and use the 
coefficient of this expansion as variational parameters
\begin{equation}
V(\mathbf{s};\boldsymbol{\alpha}) = 
\sum_{\mathbf{k}} \alpha_{\mathbf{k}} \cdot G_{\mathbf{k}}(\mathbf{s}),
\end{equation}
Given the fact that in general the FES is a rather smooth function of the CVs a 
small number of terms in this expansion will usually suffice. 
This is to be contrasted with metadynamics where a large number of Gaussians are 
used to represent $V(\mathbf{s})$. 
As we shall see below this leads to great efficiency.
In this case the gradient and the Hessian simplify 
\begin{align}
\label{gradient2}
& \frac
{\partial \Omega(\boldsymbol{\alpha})}
{\partial \alpha_{\mathbf{i}}}
= - 
\left< 
G_{\mathbf{i}}(\mathbf{s}) 
\right>_{V(\boldsymbol{\alpha})} +
\left< 
G_{\mathbf{i}}(\mathbf{s}) 
\right>_{p},
\\
& \frac
{\partial^{2} \Omega(\boldsymbol{\alpha})}
{\partial \alpha_{\mathbf{j}} \partial \alpha_{\mathbf{i}}}
 = \beta \cdot 
\mathrm{Cov}
\left[
G_{\mathbf{j}}(\mathbf{s}),
G_{\mathbf{i}}(\mathbf{s}) 
\right]_{V(\boldsymbol{\alpha})}.
\end{align}
The gradient and Hessian terms for the constant 
term $\alpha_{\mathbf{0}}$ are zero for any 
given $p(\mathbf{s})$ so we can naturally 
take the constant term as zero and drop it 
from the linear expansion.

Since the gradients and the Hessian are computed  statistically 
they are intrinsically noisy 
and one would need very long sampling times if we were to use them in 
conventional deterministic optimisation algorithms. 
Thus we turn to the vast literature on stochastic optimization 
methods~\cite{Kushner2003} and use a recent stochastic gradient descent based algorithm~\cite{Back2013a}. 
In this algorithm, we consider at iteration $n$ both the 
instantaneous iterate $\boldsymbol{\alpha}^{(n)}$ and the 
averaged iterates 
$\bar{\boldsymbol{\alpha}}^{(n)} = (n+1)^{-1} \sum_{k=0}^{n} 
\boldsymbol{\alpha}^{(k)}$. 
The instantaneous iterate is then updated using the recursion equation 
\begin{equation} 
\boldsymbol{\alpha}^{(n+1)} = \boldsymbol{\alpha}^{(n)} - \mu
\left[ 
\Omega'(\bar{\boldsymbol{\alpha}}^{(n)}) 
+ \Omega''(\bar{\boldsymbol{\alpha}}^{(n)}) 
[\boldsymbol{\alpha}^{(n)} - \bar{\boldsymbol{\alpha}}^{(n)}]
\right],
\end{equation}
where $\mu$ is a fixed step size and 
the gradient and Hessian are always obtained by 
using the averaged iterates $\bar{\boldsymbol{\alpha}}^{(n)}$, 
which amounts to taking first order Taylor expansion of the gradient 
$\Omega'(\boldsymbol{\alpha}^{(n)})$ around $\bar{\boldsymbol{\alpha}}^{(n)}$. 
As we show below, the instantaneous iterates $\boldsymbol{\alpha}^{(n)}$ 
fluctuate considerably  while their averages $\bar{\boldsymbol{\alpha}}^{(n)}$ 
 vary smoothly. 
This leads to a well behaved biasing potential 
$V(\mathbf{s};\bar{\boldsymbol{\alpha}}^{(n)})$ and to  a 
smoothly converging estimate of $F(\mathbf{s})$, either directly 
from Eq.~\ref{optimal_bias} or through reweighting 
using Eq.~\ref{US_reweighting}. 
The averaging of the iterates also allows for rather short sampling 
time at each iteration ($\sim 1$ ps in the cases examined here).
The choice of the step size is at present still a matter of trial and error 
and we expect it to depend on the system and on the functional form of  
$V(\mathbf{s};\boldsymbol{\alpha})$.
We note that in many cases it may be too costly to obtain 
the complete Hessian $\Omega''(\bar{\boldsymbol{\alpha}}^{(n)})$ so for 
practical reasons one can consider only its diagonal part as done here.
In our experience so far this does not seem to cause any ill effect. 

We now turn to  exemplifying how the new method works in practice. 
In the main text we consider only angular CVs but in fact any variable can be 
treated in a similar way (see SM). In this case the natural choice is to take  
$p(\mathbf{s})=\frac{1}{(2\pi)^d}$ where $d$ is the number of CVs biased. 
We expand $V(\mathbf{s})$ in a Fourier series,
$V(\mathbf{s}) =\sum_{\mathbf{k}} 
\alpha_{\mathbf{k}} e^{i \mathbf{k} \mathbf{s}}$,  
and use the expansion coefficients as variational parameters 
(see SM for details). 
With the chosen constant $p(\mathbf{s})$ one always 
has $\left< V(\mathbf{s}) \right>_{p} = 0$ which fixes the zero of 
$V(\mathbf{s})$ during minimization and facilitates judging the 
convergence of the simulation. 
Each calculation is started with all variational parameters set to zero, 
that is $V(\mathbf{s},\bar{\boldsymbol{\alpha}}^{(0)})=0$.

\begin{figure}[htb!]
\includegraphics[width=1.0\columnwidth]{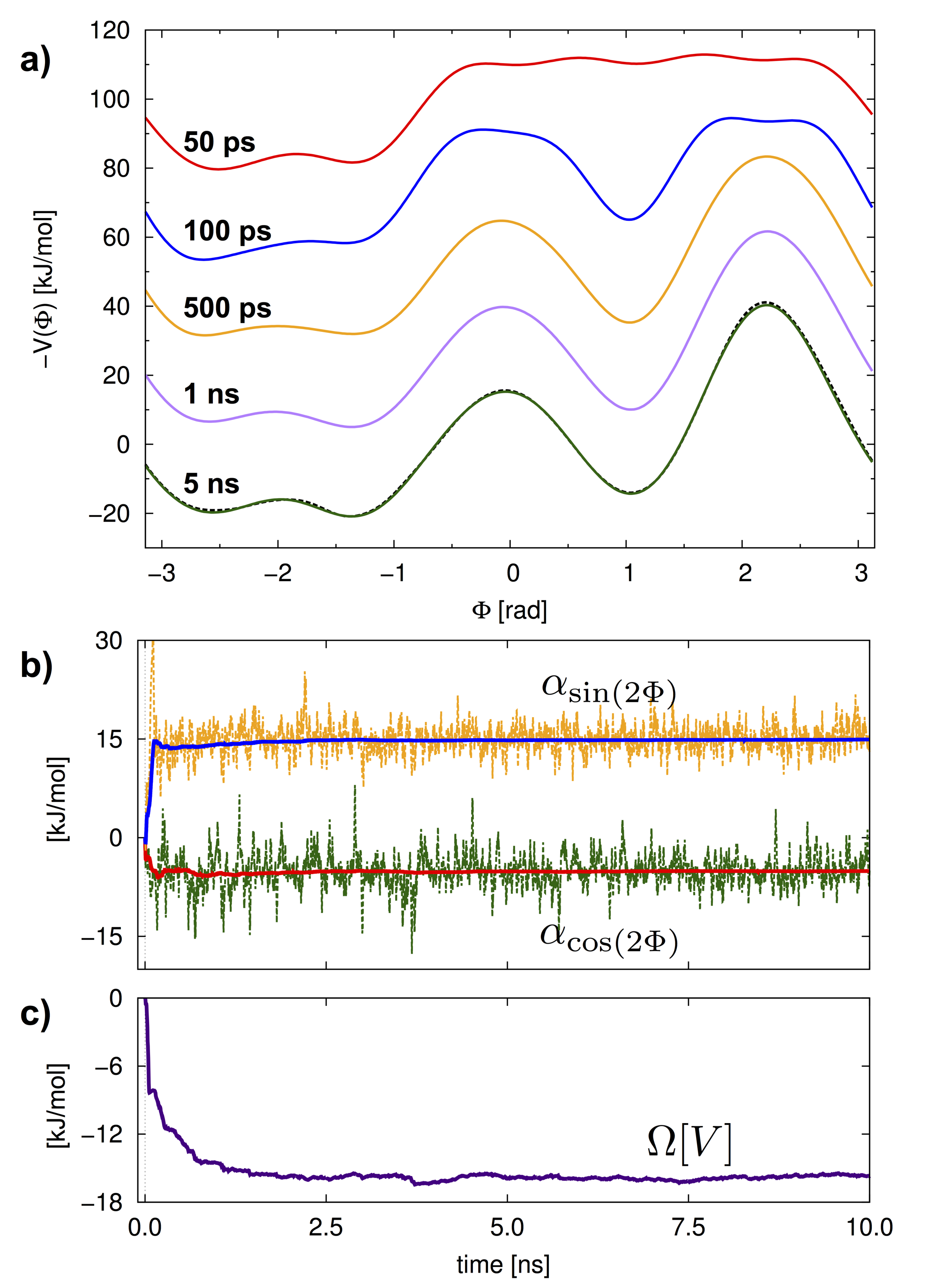}
\caption{
(Color online)
Time evolution of the bias potential during minimization for alanine dipeptide 
in vacuum at 300 K. 
Only the backbone dihedral angle $\Phi$ acts as CV and 12 basis functions are 
included in the expansion of $V(\Phi)$.
\textbf{a)} 
The $F(\Phi)$ as estimated by the negative of the bias potential 
(for clarity the potentials are shifted relative to one another by 25 kJ/mol). 
At 5 ns we also compare our variational result with a fully converged estimate 
of the FES from well-tempered metadynamics 
(black dashed line). The two curves are almost indistinguishable. 
\textbf{b)} 
Time evolution of the instantaneous and average expansion coefficients of  
$\cos(2\Phi)$ and $\sin(2\Phi)$.
\textbf{c)} 
Time evolution $\Omega[V]$ estimated by a running average from the 
start of the simulation (see SM).}
\label{Figure1}
\end{figure}

It has become customary to test any new free energy  method on 
alanine dipeptide in vacuum and we shall adhere to this tradition.  
Conventionally the FES of alanine dipeptide is described in terms of the 
two backbone dihedral angles $\Phi$ and $\Psi$ (see SM) 
but in vacuum only the 
$\Phi$ angle is a slow degree-of-freedom while $\Psi$ can be 
considered as a fast degree-of-freedom. 
Therefore, by biasing only $\Phi$ one can still 
obtain a proper sampling of phase space.
In Fig.~\ref{Figure1} we show the evolution of the minimization process when 
using only the backbone dihedral angle $\Phi$ as CV.
It is seen that in this case the bias potential $V(\Phi)$  evolves in a manner 
resembling that of standard metadynamics, 
filling progressively all the different minima
and smoothly converging to the reference free energy profile $F(\Phi)$ obtained 
with metadynamics.
In the same figure we also show two randomly chosen coefficients in the 
expansion of $ V(\Phi) $ where we observe that while their 
instantaneous values oscillates greatly their averages converge smoothly. 
The same convergent behaviour is observed in the value of $\Omega[V]$ in 
Fig.~\ref{Figure1}c. 
We have also performed a conventional calculation using both 
backbone dihedral angles $\Phi$ and $\Psi$ as CVs with similar 
satisfactory results (see SM). Solvating the alanine dipeptide 
in explicit water and using the two traditional CVs also leads to 
gratifying results (see SM).

As noted earlier in many cases the FES are rather smooth functions so one can
obtain a good  representation of $V(\mathbf{s})$  with only a minimal basis set.  
In alanine dipeptide both in vacuum and in water we obtain already a rather good 
description of the FES with only 7  basis functions per CV. 
We make use of this ability of representing the FES with a minimal basis set in 
our next example.

\begin{figure}[h]
\includegraphics[width=1.0\columnwidth]{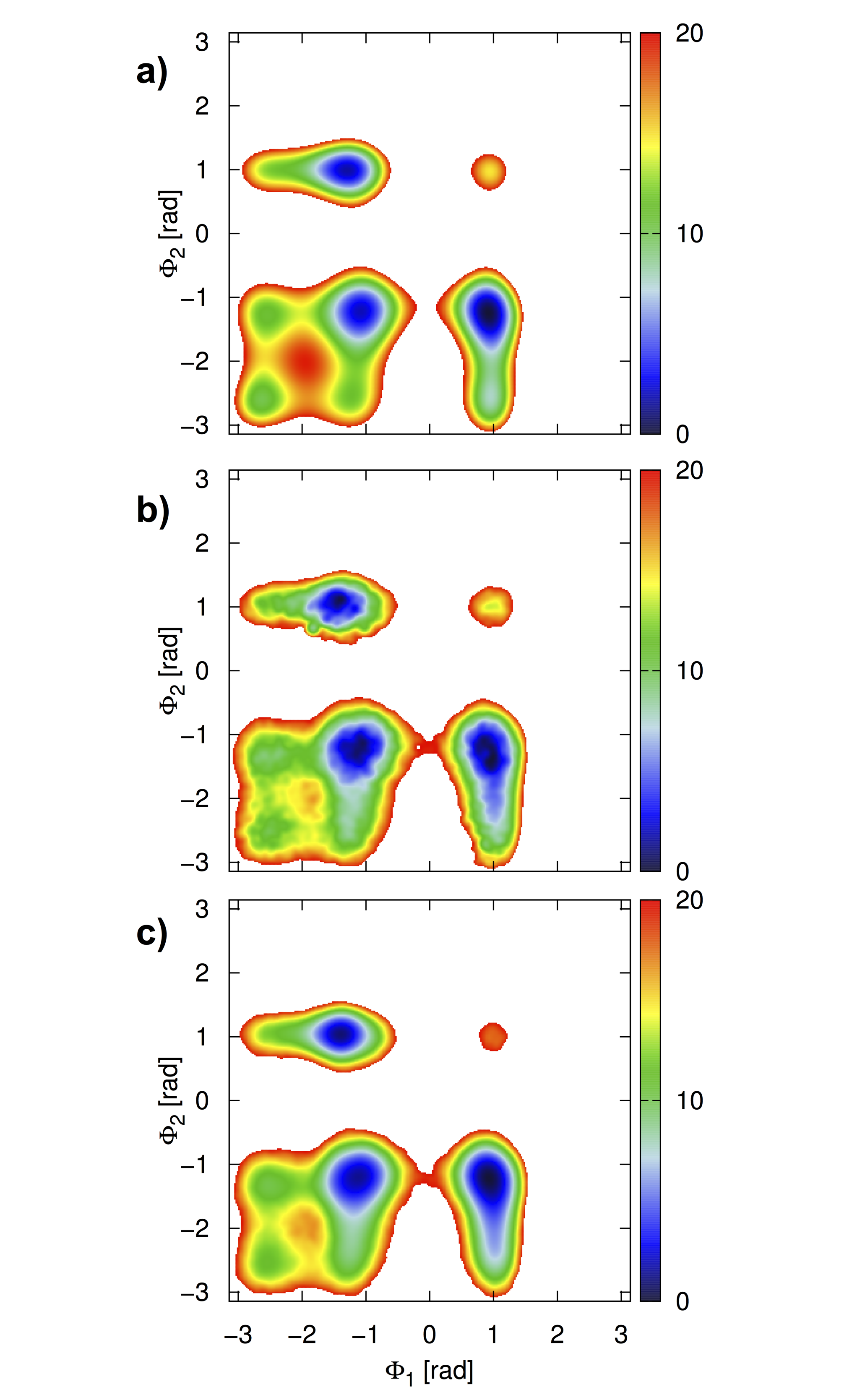}
\caption{
(Color online)
The two-dimensional FES $F(\Phi_{1},\Phi_{2})$ obtained with the 
variational approach for Ala$_{3}$ in vacuum 
at 300K using the backbone dihedral angles $\Phi_{1}$, $\Phi_{2}$, 
and $\Phi_{3}$ as CVs and 342 basis functions in the 
expansion of $V(\Phi_{1},\Phi_{2},\Phi_{3})$.
\textbf{a)} FES obtained from a projection of 
$F(\Phi_{1},\Phi_{2},\Phi_{3})$. 
\textbf{b)} FES obtained with on the fly reweighting.
\textbf{c)} Reference results from a 500 ns parallel tempering 
simulation (8 replicas with an aggregated simulation time of 4 $\mu$s). 
The color scale of the FES is given in units of kJ/mol.
All FES have their minimum value 
set to zero and are cut such that regions higher than  
8 $k_{\textrm{B}}T$ ($\approx 20$ kJ/mol) are not 
shown. See SM for further details. 
}
\label{Figure2}
\end{figure}

This is the more challenging case of a Ala$_{3}$ peptide in vacuum. 
While its conformations are  described by its six backbone dihedral angles 
$\Phi_{1}$,$\Psi_{1}$,$\Phi_{2}$,$\Psi_{2}$,$\Phi_{3}$,$\Psi_{3}$ (see SM) 
only  the three $\Phi$ angles suffice as CVs. As we increase the dimensionality 
of the CV space the number of variational parameters increases exponentially 
with $d$. 
To keep the number of variational parameters small we shall only use a 
minimal basis set. 
In the Ala$_{3}$ case this leads to use only 342 basis functions. 
With this choice during optimization all three CVs quickly become diffusive and 
the bias potential $V(\Phi_{1},\Phi_{2},\Phi_{3})$ converges 
after 50-100 ns of simulation time. 

Despite employing a minimal basis 
we get a rather good representation of the 
FES as shown in Fig.~\ref{Figure2} 
where we present a two-dimensional projection of 
$F(\Phi_{1},\Phi_{2},\Phi_{3})$ on $\Phi_{1}$ and $\Phi_{2}$ 
(projections for the other CVs can be seen in the SM). 
Any lack of full variational flexibility in the bias potential can furthermore 
be fully corrected by performing an on the fly 
reweighting during the optimization process.
As observed in Fig.~\ref{Figure2} the 
FES obtained in this manner are in excellent agreement with 
reference results from an extensive 500 ns parallel-tempering simulation. 
In the SM we show furthermore reweigthed FES for other CVs 
not biased during the simulation that also are in excellent agreement 
with the reference results.

While this approach offers a significant improvement over 
metadynamics and other similar
methods its usefulness still depends on an appropriate choice of the CVs.
Like in metadynamics, a poor choice of the CVs  will manifest itself in a 
hysteretical  behaviour during the optimization process (see SM). 
However our variational approach with its potential for  handling many CVs can 
greatly alleviate the problem. 
A further help in this direction is the possibility of adding  variational 
flexibility in the definition of the CVs.

To improve upon the method we can borrow all the ideas that have been applied 
to metadynamics like parallel tempering~\cite{Bussi2006_PTMetaD}, 
multiple walkers~\cite{Raiteri2006}, or bias-exchange~\cite{Piana2007}. 
Furthermore, metadynamics itself can be used to sample the averages needed in 
Eqs.~\ref{gradient1} and~\ref{hessian1} 
by employing a recent improved reweighting scheme~\cite{Tiwary2014}. 
The sampling power of the variational approach can thus be further enhanced by 
biasing the metadynamics with CVs different from  $\mathbf{s}$. 

The main result of this paper is the introduction of the functional $\Omega [V]$ 
and the practical demonstration of its usefulness. 
We believe that there is ample room for improvement. 
The optimization procedure presented here is not necessarily optimal 
and different systems and CVs might require different optimization strategies 
and different basis set.
We plan to explore a number of alternative procedures.
For instance one could think of setting up an iterative procedure 
in which  an approximate calculation is made for $F_{0}(\mathbf{s})$ 
using Eq.~\ref{optimal_bias} at the early stages of the calculation. 
One can then insert into Eq.~\ref{omega1} a new $ p_{0} 
(\mathbf{s})=\frac{e^{-\beta 
F_{0}(\mathbf{s})}}{\int d\mathbf{s} \, e^{-\beta 
F_{0}(\mathbf{s})}}$. The resulting functional is then optimized  
and the procedure iterated until at convergence after $k$ steps
$ p_{k} (\mathbf{s}) = \frac{e^{-\beta F_{k}(\mathbf{s})}}{\int d\mathbf{s} \,
e^{-\beta F_{k}(\mathbf{s})}}$  
and $V_{k}(\mathbf{s}) \approx 0$ to the desired accuracy.

These brief discussion on the potential for improvements and modifications of 
the scheme is by far not exhaustive but is meant to indicate some of the future 
lines of investigation and indicate the in this very first application we are 
only using a small fraction of the potentialities of this 
method and that much more exciting developments are to be expected. 
We would also like to point out the potential of our method in the development of a more 
rigorous coarse graining procedure.

Finally we note that the systems considered here are by necessity simple, as 
conventionally done when introducing a completely new method. The strengths 
and limitations of our approach will become clearer as it is further developed.

The method has been implemented in a development version of the 
PLUMED 2~\cite{Tribello2014} plug-in 
and will be made publicly available in the coming future.

\begin{acknowledgments}
The authors would like to thank David Chandler for insightful discussions.
All calculation were performed on the Brutus HPC cluster at ETH Zurich.
We acknowledge the European Union grant ERC-2009-AdG-247075 for funding.
\end{acknowledgments}

\bibliography{Bibliography}

\begin{thebibliography}{19}%
\makeatletter
\providecommand \@ifxundefined [1]{%
 \@ifx{#1\undefined}
}%
\providecommand \@ifnum [1]{%
 \ifnum #1\expandafter \@firstoftwo
 \else \expandafter \@secondoftwo
 \fi
}%
\providecommand \@ifx [1]{%
 \ifx #1\expandafter \@firstoftwo
 \else \expandafter \@secondoftwo
 \fi
}%
\providecommand \natexlab [1]{#1}%
\providecommand \enquote  [1]{``#1''}%
\providecommand \bibnamefont  [1]{#1}%
\providecommand \bibfnamefont [1]{#1}%
\providecommand \citenamefont [1]{#1}%
\providecommand \href@noop [0]{\@secondoftwo}%
\providecommand \href [0]{\begingroup \@sanitize@url \@href}%
\providecommand \@href[1]{\@@startlink{#1}\@@href}%
\providecommand \@@href[1]{\endgroup#1\@@endlink}%
\providecommand \@sanitize@url [0]{\catcode `\\12\catcode `\$12\catcode
  `\&12\catcode `\#12\catcode `\^12\catcode `\_12\catcode `\%12\relax}%
\providecommand \@@startlink[1]{}%
\providecommand \@@endlink[0]{}%
\providecommand \url  [0]{\begingroup\@sanitize@url \@url }%
\providecommand \@url [1]{\endgroup\@href {#1}{\urlprefix }}%
\providecommand \urlprefix  [0]{URL }%
\providecommand \Eprint [0]{\href }%
\providecommand \doibase [0]{http://dx.doi.org/}%
\providecommand \selectlanguage [0]{\@gobble}%
\providecommand \bibinfo  [0]{\@secondoftwo}%
\providecommand \bibfield  [0]{\@secondoftwo}%
\providecommand \translation [1]{[#1]}%
\providecommand \BibitemOpen [0]{}%
\providecommand \bibitemStop [0]{}%
\providecommand \bibitemNoStop [0]{.\EOS\space}%
\providecommand \EOS [0]{\spacefactor3000\relax}%
\providecommand \BibitemShut  [1]{\csname bibitem#1\endcsname}%
\let\auto@bib@innerbib\@empty
\bibitem [{\citenamefont {Torrie}\ and\ \citenamefont
  {Valleau}(1977)}]{Torrie1977a}%
  \BibitemOpen
  \bibfield  {author} {\bibinfo {author} {\bibfnamefont {G.}~\bibnamefont
  {Torrie}}\ and\ \bibinfo {author} {\bibfnamefont {J.}~\bibnamefont
  {Valleau}},\ }\href {\doibase 10.1016/0021-9991(77)90121-8} {\bibfield
  {journal} {\bibinfo  {journal} {J. Comput. Phys.}\ }\textbf {\bibinfo
  {volume} {23}},\ \bibinfo {pages} {187} (\bibinfo {year} {1977})}\BibitemShut
  {NoStop}%
\bibitem [{\citenamefont {Huber}\ \emph {et~al.}(1994)\citenamefont {Huber},
  \citenamefont {Torda},\ and\ \citenamefont {Gunsteren}}]{Huber1994}%
  \BibitemOpen
  \bibfield  {author} {\bibinfo {author} {\bibfnamefont {T.}~\bibnamefont
  {Huber}}, \bibinfo {author} {\bibfnamefont {A.~E.}\ \bibnamefont {Torda}}, \
  and\ \bibinfo {author} {\bibfnamefont {W.~F.}\ \bibnamefont {Gunsteren}},\
  }\href {\doibase 10.1007/bf00124016} {\bibfield  {journal} {\bibinfo
  {journal} {J Computer-Aided Mol Des}\ }\textbf {\bibinfo {volume} {8}},\
  \bibinfo {pages} {695} (\bibinfo {year} {1994})}\BibitemShut {NoStop}%
\bibitem [{\citenamefont {Darve}\ and\ \citenamefont
  {Pohorille}(2001)}]{Darve2001}%
  \BibitemOpen
  \bibfield  {author} {\bibinfo {author} {\bibfnamefont {E.}~\bibnamefont
  {Darve}}\ and\ \bibinfo {author} {\bibfnamefont {A.}~\bibnamefont
  {Pohorille}},\ }\href {\doibase 10.1063/1.1410978} {\bibfield  {journal}
  {\bibinfo  {journal} {J. Chem. Phys.}\ }\textbf {\bibinfo {volume} {115}},\
  \bibinfo {pages} {9169} (\bibinfo {year} {2001})}\BibitemShut {NoStop}%
\bibitem [{\citenamefont {Wang}\ and\ \citenamefont
  {Landau}(2001)}]{WangLandau2001a}%
  \BibitemOpen
  \bibfield  {author} {\bibinfo {author} {\bibfnamefont {F.}~\bibnamefont
  {Wang}}\ and\ \bibinfo {author} {\bibfnamefont {D.}~\bibnamefont {Landau}},\
  }\href {\doibase 10.1103/PhysRevLett.86.2050} {\bibfield  {journal} {\bibinfo
   {journal} {Phys. Rev. Lett.}\ }\textbf {\bibinfo {volume} {86}},\ \bibinfo
  {pages} {2050} (\bibinfo {year} {2001})}\BibitemShut {NoStop}%
\bibitem [{\citenamefont {Laio}\ and\ \citenamefont
  {Parrinello}(2002)}]{Laio2002a}%
  \BibitemOpen
  \bibfield  {author} {\bibinfo {author} {\bibfnamefont {A.}~\bibnamefont
  {Laio}}\ and\ \bibinfo {author} {\bibfnamefont {M.}~\bibnamefont
  {Parrinello}},\ }\href {\doibase 10.1073/pnas.202427399} {\bibfield
  {journal} {\bibinfo  {journal} {Proc. Natl. Acad. Sci. U.S.A.}\ }\textbf
  {\bibinfo {volume} {99}},\ \bibinfo {pages} {12562} (\bibinfo {year}
  {2002})}\BibitemShut {NoStop}%
\bibitem [{\citenamefont {Hansmann}\ and\ \citenamefont
  {Wille}(2002)}]{Hansmann2002a}%
  \BibitemOpen
  \bibfield  {author} {\bibinfo {author} {\bibfnamefont {U.}~\bibnamefont
  {Hansmann}}\ and\ \bibinfo {author} {\bibfnamefont {L.}~\bibnamefont
  {Wille}},\ }\href {http://dx.doi.org/10.1103/PhysRevLett.88.068105}
  {\bibfield  {journal} {\bibinfo  {journal} {Phys. Rev. Lett.}\ }\textbf
  {\bibinfo {volume} {88}} (\bibinfo {year} {2002})}\BibitemShut {NoStop}%
\bibitem [{\citenamefont {Maragliano}\ and\ \citenamefont
  {Vanden-Eijnden}(2006)}]{Maragliano2006a}%
  \BibitemOpen
  \bibfield  {author} {\bibinfo {author} {\bibfnamefont {L.}~\bibnamefont
  {Maragliano}}\ and\ \bibinfo {author} {\bibfnamefont {E.}~\bibnamefont
  {Vanden-Eijnden}},\ }\href {\doibase 10.1016/j.cplett.2006.05.062} {\bibfield
   {journal} {\bibinfo  {journal} {Chem. Phys. Lett.}\ }\textbf {\bibinfo
  {volume} {426}},\ \bibinfo {pages} {168} (\bibinfo {year}
  {2006})}\BibitemShut {NoStop}%
\bibitem [{\citenamefont {Abrams}\ and\ \citenamefont
  {Vanden-Eijnden}(2010)}]{Abrams2010a}%
  \BibitemOpen
  \bibfield  {author} {\bibinfo {author} {\bibfnamefont {C.~F.}\ \bibnamefont
  {Abrams}}\ and\ \bibinfo {author} {\bibfnamefont {E.}~\bibnamefont
  {Vanden-Eijnden}},\ }\href {\doibase 10.1073/pnas.0914540107} {\bibfield
  {journal} {\bibinfo  {journal} {Proc. Natl. Acad. Sci. U.S.A.}\ }\textbf
  {\bibinfo {volume} {107}},\ \bibinfo {pages} {4961} (\bibinfo {year}
  {2010})}\BibitemShut {NoStop}%
\bibitem [{\citenamefont {Maragakis}\ \emph {et~al.}(2009)\citenamefont
  {Maragakis}, \citenamefont {van~der Vaart},\ and\ \citenamefont
  {Karplus}}]{Maragakis2009a}%
  \BibitemOpen
  \bibfield  {author} {\bibinfo {author} {\bibfnamefont {P.}~\bibnamefont
  {Maragakis}}, \bibinfo {author} {\bibfnamefont {A.}~\bibnamefont {van~der
  Vaart}}, \ and\ \bibinfo {author} {\bibfnamefont {M.}~\bibnamefont
  {Karplus}},\ }\href {\doibase 10.1021/jp808381s} {\bibfield  {journal}
  {\bibinfo  {journal} {J. Phys. Chem. B}\ }\textbf {\bibinfo {volume} {113}},\
  \bibinfo {pages} {4664} (\bibinfo {year} {2009})}\BibitemShut {NoStop}%
\bibitem [{\citenamefont {Barducci}\ \emph {et~al.}(2011)\citenamefont
  {Barducci}, \citenamefont {Bonomi},\ and\ \citenamefont
  {Parrinello}}]{Barducci2011a}%
  \BibitemOpen
  \bibfield  {author} {\bibinfo {author} {\bibfnamefont {A.}~\bibnamefont
  {Barducci}}, \bibinfo {author} {\bibfnamefont {M.}~\bibnamefont {Bonomi}}, \
  and\ \bibinfo {author} {\bibfnamefont {M.}~\bibnamefont {Parrinello}},\
  }\href {\doibase 10.1002/.31} {\bibfield  {journal} {\bibinfo  {journal}
  {WIREs: Comp. Mol. Sci.}\ }\textbf {\bibinfo {volume} {1}},\ \bibinfo {pages}
  {826} (\bibinfo {year} {2011})}\BibitemShut {NoStop}%
\bibitem [{\citenamefont {Barducci}\ \emph {et~al.}(2008)\citenamefont
  {Barducci}, \citenamefont {Bussi},\ and\ \citenamefont
  {Parrinello}}]{Barducci2008_WTMetaD}%
  \BibitemOpen
  \bibfield  {author} {\bibinfo {author} {\bibfnamefont {A.}~\bibnamefont
  {Barducci}}, \bibinfo {author} {\bibfnamefont {G.}~\bibnamefont {Bussi}}, \
  and\ \bibinfo {author} {\bibfnamefont {M.}~\bibnamefont {Parrinello}},\
  }\href@noop {} {\bibfield  {journal} {\bibinfo  {journal} {Phys. Rev. Lett.}\
  }\textbf {\bibinfo {volume} {100}},\ \bibinfo {pages} {020603} (\bibinfo
  {year} {2008})}\BibitemShut {NoStop}%
\bibitem [{\citenamefont {Dama}\ \emph {et~al.}(2014)\citenamefont {Dama},
  \citenamefont {Parrinello},\ and\ \citenamefont {Voth}}]{Dama2014a}%
  \BibitemOpen
  \bibfield  {author} {\bibinfo {author} {\bibfnamefont {J.~F.}\ \bibnamefont
  {Dama}}, \bibinfo {author} {\bibfnamefont {M.}~\bibnamefont {Parrinello}}, \
  and\ \bibinfo {author} {\bibfnamefont {G.~A.}\ \bibnamefont {Voth}},\
  }\href@noop {} {\bibfield  {journal} {\bibinfo  {journal} {Phys. Rev. Lett.}\
  }\textbf {\bibinfo {volume} {112}} (\bibinfo {year} {2014})}\BibitemShut
  {NoStop}%
\bibitem [{\citenamefont {Kushner}\ and\ \citenamefont
  {Yin}(2003)}]{Kushner2003}%
  \BibitemOpen
  \bibfield  {author} {\bibinfo {author} {\bibfnamefont {H.~J.}\ \bibnamefont
  {Kushner}}\ and\ \bibinfo {author} {\bibfnamefont {G.~G.}\ \bibnamefont
  {Yin}},\ }\href {\doibase 10.1007/b97441} {\emph {\bibinfo {title}
  {Stochastic Approximation and Recursive Algorithms and Applications}}}\
  (\bibinfo  {publisher} {Springer-Verlag},\ \bibinfo {year}
  {2003})\BibitemShut {NoStop}%
\bibitem [{\citenamefont {Bach}\ and\ \citenamefont
  {Moulines}(2013)}]{Back2013a}%
  \BibitemOpen
  \bibfield  {author} {\bibinfo {author} {\bibfnamefont {F.}~\bibnamefont
  {Bach}}\ and\ \bibinfo {author} {\bibfnamefont {E.}~\bibnamefont
  {Moulines}},\ }in\ \href@noop {} {\emph {\bibinfo {booktitle} {Advances in
  Neural Information Processing Systems 26}}},\ \bibinfo {editor} {edited by\
  \bibinfo {editor} {\bibfnamefont {C.}~\bibnamefont {Burges}}, \bibinfo
  {editor} {\bibfnamefont {L.}~\bibnamefont {Bottou}}, \bibinfo {editor}
  {\bibfnamefont {M.}~\bibnamefont {Welling}}, \bibinfo {editor} {\bibfnamefont
  {Z.}~\bibnamefont {Ghahramani}}, \ and\ \bibinfo {editor} {\bibfnamefont
  {K.}~\bibnamefont {Weinberger}}}\ (\bibinfo  {publisher} {Curran Associates,
  Inc.},\ \bibinfo {year} {2013})\ pp.\ \bibinfo {pages} {773--781}\BibitemShut
  {NoStop}%
\bibitem [{\citenamefont {Bussi}\ \emph {et~al.}(2006)\citenamefont {Bussi},
  \citenamefont {Gervasio}, \citenamefont {Laio},\ and\ \citenamefont
  {Parrinello}}]{Bussi2006_PTMetaD}%
  \BibitemOpen
  \bibfield  {author} {\bibinfo {author} {\bibfnamefont {G.}~\bibnamefont
  {Bussi}}, \bibinfo {author} {\bibfnamefont {F.~L.}\ \bibnamefont {Gervasio}},
  \bibinfo {author} {\bibfnamefont {A.}~\bibnamefont {Laio}}, \ and\ \bibinfo
  {author} {\bibfnamefont {M.}~\bibnamefont {Parrinello}},\ }\href {\doibase
  10.1021/ja062463w} {\bibfield  {journal} {\bibinfo  {journal} {J. Am. Chem.
  Soc.}\ }\textbf {\bibinfo {volume} {128}},\ \bibinfo {pages} {13435}
  (\bibinfo {year} {2006})}\BibitemShut {NoStop}%
\bibitem [{\citenamefont {Raiteri}\ \emph {et~al.}(2006)\citenamefont
  {Raiteri}, \citenamefont {Laio}, \citenamefont {Gervasio}, \citenamefont
  {Micheletti},\ and\ \citenamefont {Parrinello}}]{Raiteri2006}%
  \BibitemOpen
  \bibfield  {author} {\bibinfo {author} {\bibfnamefont {P.}~\bibnamefont
  {Raiteri}}, \bibinfo {author} {\bibfnamefont {A.}~\bibnamefont {Laio}},
  \bibinfo {author} {\bibfnamefont {F.~L.}\ \bibnamefont {Gervasio}}, \bibinfo
  {author} {\bibfnamefont {C.}~\bibnamefont {Micheletti}}, \ and\ \bibinfo
  {author} {\bibfnamefont {M.}~\bibnamefont {Parrinello}},\ }\href {\doibase
  10.1021/jp054359r} {\bibfield  {journal} {\bibinfo  {journal} {J. Phys. Chem.
  B}\ }\textbf {\bibinfo {volume} {110}},\ \bibinfo {pages} {3533} (\bibinfo
  {year} {2006})}\BibitemShut {NoStop}%
\bibitem [{\citenamefont {Piana}\ and\ \citenamefont {Laio}(2007)}]{Piana2007}%
  \BibitemOpen
  \bibfield  {author} {\bibinfo {author} {\bibfnamefont {S.}~\bibnamefont
  {Piana}}\ and\ \bibinfo {author} {\bibfnamefont {A.}~\bibnamefont {Laio}},\
  }\href {\doibase 10.1021/jp067873l} {\bibfield  {journal} {\bibinfo
  {journal} {J. Phys. Chem. B}\ }\textbf {\bibinfo {volume} {111}},\ \bibinfo
  {pages} {4553} (\bibinfo {year} {2007})}\BibitemShut {NoStop}%
\bibitem [{\citenamefont {Tiwary}\ and\ \citenamefont
  {Parrinello}(2014)}]{Tiwary2014}%
  \BibitemOpen
  \bibfield  {author} {\bibinfo {author} {\bibfnamefont {P.}~\bibnamefont
  {Tiwary}}\ and\ \bibinfo {author} {\bibfnamefont {M.}~\bibnamefont
  {Parrinello}},\ }\href@noop {} {\bibfield  {journal} {\bibinfo  {journal} {J.
  Phys. Chem. B}\ }\textbf {\bibinfo {volume} {DOI: 10.1021/jp504920s}}
  (\bibinfo {year} {2014})}\BibitemShut {NoStop}%
\bibitem [{\citenamefont {Tribello}\ \emph {et~al.}(2014)\citenamefont
  {Tribello}, \citenamefont {Bonomi}, \citenamefont {Branduardi}, \citenamefont
  {Camilloni},\ and\ \citenamefont {Bussi}}]{Tribello2014}%
  \BibitemOpen
  \bibfield  {author} {\bibinfo {author} {\bibfnamefont {G.~A.}\ \bibnamefont
  {Tribello}}, \bibinfo {author} {\bibfnamefont {M.}~\bibnamefont {Bonomi}},
  \bibinfo {author} {\bibfnamefont {D.}~\bibnamefont {Branduardi}}, \bibinfo
  {author} {\bibfnamefont {C.}~\bibnamefont {Camilloni}}, \ and\ \bibinfo
  {author} {\bibfnamefont {G.}~\bibnamefont {Bussi}},\ }\href
  {http://dx.doi.org/10.1016/j.cpc.2013.09.018} {\bibfield  {journal} {\bibinfo
   {journal} {Comput. Phys. Commun.}\ }\textbf {\bibinfo {volume} {185}},\
  \bibinfo {pages} {604} (\bibinfo {year} {2014})}\BibitemShut {NoStop}%
\end{thebibliography}%

\end{document}